\documentstyle[prd,aps,epsf,floats,amsfonts,amssymb,amsmath]{revtex}

\begin{document}
 \draft
  \title{Pulsar Kicks Induced by Spin Flavor Oscillations of Neutrinos
  in Gravitational Fields}
%  \author{xxx}
\author{G. Lambiase$^{a,b}$\thanks{E-mail: lambiase@sa.infn.it}}
%%, G. Papini$^c$\thanks{E-mail:papini@uregina.ca}, and G.
%%Scarpetta$^{a,b}$\thanks{scarpetta@sa.infn.it}}
\address{$^a$Dipartimento di Fisica "E.R. Caianiello",
 Universit\'a di Salerno, 84081 Baronissi (Sa), Italy.}
  \address{$^b$INFN - Gruppo Collegato di Salerno, Italy.}
%  \address{$^c xxx$}
\date{\today}
\maketitle
\begin{abstract}
The origin of pulsar kicks is reviewed in the framework of the
spin-flip conversion of neutrinos propagating in the gravitational
field of a magnetized protoneutron star. We find that for a mass
in rotation with angular velocity ${\bbox \omega}$, the spin
connections entering in the Dirac equation give rise to the
coupling term ${\bbox \omega}\cdot {\bf p}$, being ${\bf p}$ the
neutrino momentum. Such a coupling can be responsible of pulsar
kicks owing to the neutrino emission asymmetry generated by the
relative orientation of ${\bf p}$ with respect to ${\bbox
\omega}$. For our estimations, the large non standard neutrino
magnetic momentum, $\mu_\nu \lesssim 10^{-11}\mu_B$, is
considered.

%Besides, typical values of parameters characterizing pulsar properties are
%%%lead to values of neutrino oscillation parameters $\Delta m^2 - \sin^2\theta$ ($\Delta m^2$
%%%is the mass squared difference and $\theta$ the vacuum mixing angle)
%compatible with the best fit of the existing data on solar and
%atmospheric neutrino oscillations.
\end{abstract}
% \Keyword(s){Neutrino Mass and Mixing, QFT in curved spacetime, Pulsar}
%\vskip2pc]

\section{Introduction}

The origin of the pulsar velocity represents till now an unsolved
and still discussed issue of the modern astrophysics. Observations
show that pulsars have a very high proper motion (velocity) with
respect to the surrounding stars, with a three-dimensional
galactic speed of $450\pm 90$Km/sec up to values greater than
1000Km/sec \cite{Lyne,arzoumanian,corber,hansen,fryer,janka}. This
suggests that nascent pulsars undergo to some kind of impulse
(kick). After the supernova collapse of a massive star, neutrinos
carry away almost all ($99\%$) the gravitational binding energy
($3\times 10^{53}$erg). The momentum taken by them is about
$10^{43}$gr cm/sec. An anisotropy of $\sim 1\%$ of the momenta
distribution of the outgoing neutrinos would suffice to account
for the neutron star recoil of 300Km/sec. Even though many
mechanisms have been proposed, the origin of the pulsar kick
remains an open issue.

An interesting and elegant mechanism to generate the pulsar
velocity, which relies on the neutrino oscillation physics in
presence of an intense magnetic field, has been proposed by
Kusenko and Segr\'e (KS) \cite{kusenkoPRL}. The basic idea is the
following. Electron neutrinos $\nu_e$ are emitted from a surface
placed at a distance from the center of a pulsar greater than the
surfaces corresponding to muon neutrinos $\nu_\mu$ and tau
neutrinos $\nu_\tau$. Such surfaces are called {\it
neutrinospheres}. Under suitable conditions, a resonant
oscillation $\nu_e\to \nu_{\mu,\tau}$ may occur between the
$\nu_e$ and $\nu_{\mu,\tau}$ neutrinospheres. The neutrinos
$\nu_e$ are trapped by the medium (due to neutral and charged
interactions) but neutrinos $\nu_{\mu,\tau}$ generate via
oscillations can escape from the protoneutron star being outside
of their neutrinosphere. Thus, the "surface of the resonance" acts
as an "effective muon/tau neutrinosphere". In the presence of a
magnetic field (or some other nonisotropic effect), the surface of
resonance can be distorted and the energy flux turns out to be
generated anisotropically. In \cite{kusenkoPRL}, the anisotropy of
the neutrino emission is {\it driven} by the polarization of the
medium due to the magnetic field ${\bf B}$. The usual MSW
resonance conditions turn out to be modified by the term
\cite{pal,DOli,capone,DoliPLB,elmfors}
 \[
\frac{eG_F}{\sqrt{2}}\sqrt[3]{\frac{3n_e}{\pi^4}}\,\,{\bf B}\cdot
{\hat {\bf p}}\,,
 \]
where ${\hat {\bf p}}={\bf p}/p$, ${\bf p}$ is the neutrino
momentum, $e$ is the electric charge, $G_F$ is the Fermi constant,
and $n_e$ is the electron density. Nevertheless, the neutrino
masses required for the KS mechanism seem to be inconsistent with
the present limits on the masses of standard electroweak
neutrinos. These limits do not apply to sterile neutrinos (they
may have only a small mixing angle with the ordinary neutrinos)
\cite{kusenkoPLB,fuller}. Papers dealing with the origin of pulsar
kicks can be found in
\cite{kusenkoPRD99,AKM,barkovich,bark,burrows,chandra,chugai,cuesta,dorofeev,duncanI,elizalde,erdas,goyal,grasso,gott,harrison,horvat,JankaPRD,laiqian,lambiasePiombino,lambiasePRD,loveridge,nardi,qian,voloshin}.

The aim of this paper is to suggest a further mechanism to
generate pulsar kicks. It is based on the spin flavor conversion
of neutrinos propagating in a gravitational field generated by a
rotating source. Even though the gravitational field per se cannot
induce neutrino oscillations, unless a violation of the
equivalence principle is invoked \cite{gasp-osc,halprin}, it
affects the resonance conditions (hence the probability that
left-handed neutrinos convert into right handed neutrinos, the
latter being sterile may escape from the protoneutron star). Such
a modification is induced by spin connections entering in the
Dirac equation in curved spacetimes. They give rise to the
coupling term $\sim {\bbox \omega}\cdot {\bf p}$, where ${\bbox
\omega}$ is the angular velocity of the gravitational source. The
relative orientation of neutrino momenta with respect to the
angular velocity determines an asymmetry of the neutrino emission,
hence it may generate the pulsar kicks.

The paper is organized as follows. In Sect. 2 we shortly review
the Dirac equation in curved space-times. Sect. 3 is devoted to
the computation of the fractional asymmetry. The resonance and
adiabatic conditions, as well as the spin flip probability are
also discussed. Conclusions are drawn in Sect. 4.

\section{Dirac Equation in the Lense-Thirring Geometry}

The phase of neutrinos propagating in a curved background is
generalized as
\cite{CAR,ahluwalia,alsing,bhattacharya,capozziello,formengo,konno,lambiaseEPJ,papini,pereira,wudka}
\begin{equation}\label{1}
\vert\psi_{f}(\lambda)\rangle = \sum_{j} U_{f j}\,
e^{i\int_{\lambda_0}^{\lambda}P\cdot
p_{null}d\lambda^{\prime}}\vert\nu_j\rangle\,{,}
\end{equation}
where $f$ is the flavor index and $j$ the mass index. $U_{f j}$
are the matrix elements transforming flavor and mass bases.
$P\cdot p_{null}=P_\mu p^\mu_{null}$, where $P_\mu$ is the
four--momentum operator generating space--time translation of the
eigenstates and $p^{\mu}_{null}=dx^{\mu}/d\lambda$ is the tangent
vector to the neutrino world-line $x^{\mu}$, parameterized by
$\lambda$. The covariant Dirac equation in curved space--time is
(in natural units) \cite{WEI} $
[i\gamma^{\mu}(x)D_{\mu}-m]\psi=0$, where the matrices
$\gamma^{\mu}(x)$ are related to the usual Dirac matrices
$\gamma^{\hat{a}}$ by means of the vierbein fields
$e^{\mu}_{\hat{a}}(x)$, i.e.
$\gamma^{\mu}(x)=e^{\mu}_{\hat{a}}(x)\gamma^{\hat{a}}$. The Greek
(Latin with hat) indices refer to curved (flat) space--time.
$D_{\mu}$ is defined as $D_{\mu}=\partial_{\mu}+\Gamma_{\mu}(x)$,
where $\Gamma_{\mu}(x)$ are the spin connections
$\Gamma_{\mu}(x)=\frac{1}{8} [\gamma^{\hat{a}},
\gamma^{\hat{b}}]e^{\nu}_{\hat{a}}e_{\nu
 \hat{b};\mu}$ (semicolon represents the covariant derivative).
After some manipulations, the spin connections can be cast in the
form \cite{CAR}
 \begin{equation}\label{AG0}
\gamma^\mu\Gamma_{\mu}=\gamma^{\hat{a}}e^{\mu}_{\hat{a}}
\left\{iA_{G\mu}\left[-(-g)^{-1/2}\, \gamma^5\right]\right\}\,,
\end{equation}
where
\begin{equation}\label{AG}
A_G^{\mu}=\frac{1}{4}\sqrt{-g}e^{\mu}_{\hat{a}}\varepsilon^{\hat{d}\hat{a}\hat{b}\hat{c}}
(e_{\hat{b}\nu;\sigma}-e_{\hat{b}\sigma;\nu})e^{\nu}_{\hat{c}}e^{\sigma}_{\hat{d}}\,,
\end{equation}
$g=det(g_{\mu\nu})$. For geometries with a spherical symmetry,
such as the Schwarzschild space-time, $A_{G}^{\mu}$ vanishes. For
rotating gravitational sources, instead, $A_G^\mu$ acquires non
vanishing components. To prove this, let us consider the
Lense-Thirring geometry, i.e. the geometry of a rotating source.
The line element (in the weak field approximation) is given by
 \begin{equation}\label{9}
 ds^2=(1-\phi)(dt)^2-(1+\phi)(d{\bf x})^2
-2{\bf h}\cdot d{\bf x}dt \,,
 \end{equation}
where
 \[
{\bf x}=(x, y, z)\,, \quad \phi=\frac{2GM}{r}\,, \quad {\bf
h}=\frac{4GMR^2}{5r^3}\,{\bbox \omega}\wedge {\bf x}\,,
 \]
$M$ is the gravitational mass, and $R$ its radius. The vierbein
fields associated to the metric (\ref{9}) are
 \[
 e_{0\, {\hat 0}}=(1+\phi)\,, \quad e_{0\, {\hat i}}=-h^i\,, \quad  e_{j\,{\hat
i}}=-(1-\phi)\delta_{ji}\,,
 \]
From Eq. (\ref{AG}) one finds
 \[
A_G^\mu(x)=\left(0, -\frac{4}{5}\frac{GM R^2}{r^3}\, {\bbox
\omega}^\prime\right)\,,
 \]
where
\[
 {\bbox \omega}^\prime={\bbox \omega}-\frac{3({\bbox \omega}\cdot {\bf x })\, {\bf x}}{r^2}\,.
 \]
%The non vanishing $A_G^\mu$ is an indication of a preferred
%direction related to the angular velocity of the source.

The operator $\gamma^\mu\Gamma_\mu\sim\gamma^\mu A_{G\,\mu}
\gamma^5$ acts differently on left- and right-handed neutrino
states \cite{CAR}. By writing $\gamma^5={\cal P}_R-{\cal P}_L$,
where ${\cal P}_{L,R}=(1\mp\gamma^5)/2$, one immediately sees that
left- and right-handed neutrinos acquire a gravitational potential
which is opposite for the two helicities. The dispersion relation
of neutrinos turns out to be modified by the term \cite{CAR}
$p^\mu A_{G\, \mu}\gamma^5=-{\bf p}\cdot {\bf A}_G\gamma^5$. In
the case of neutrino oscillations, it is convenient to add,
without physical consequences, a term proportional to the identity
matrix, so that $\gamma^5$ can be replaced by the left-handed
projection operator ${\cal P}_L$. Spin-gravity coupling terms can
be then pushed in the left-handed sector of the neutrinos
effective Hamiltonian (see Eqs. (\ref{11})-(\ref{12b})).

\section{The Asymmetric Emission of Neutrinos From Protoneutron Stars}

Neutrinos inside their neutrinospheres are trapped by the weak
interactions with the background matter. They therefore acquire
the potential energy
 \[
 V_{\nu_{f}}\simeq 3.8\, \frac{\rho}{10^{14}\mbox{gr cm}^{-3}}\, y_{f}(r,t)\,
 \mbox{eV}\,,
 \]
where $f=e, \mu, \tau$, $\rho$ is the matter density,
$y_e=Y_e-1/3$ and $y_{\mu, \tau}=Y_e-1$. In these expressions,
$Y_e$ is the electron fraction. In the present paper we shall
consider the case in which matter induced effective potential is
$|V_{\nu_e}|\ll 1$ (as shown in \cite{valle}, $V_{\nu_e}$ may
cross the zero at $r\sim 12$km). This occurs in the regions where
the electron fraction $Y_e$ assumes the value $\approx 1/3$
($y_e\ll 1 $)
\cite{athar,pons,mcLaughlin,peltoniemi,caldwell,peres,voloshin,valle}.

Protoneutron stars posses large magnetic fields which may vary in
a range of several order of magnitude. In such astrophysical
systems, the interaction of neutrinos with (uniform) magnetic
fields is described by the Lagrangian \cite{okun,muSM,haxton}
 \[
    {\cal L}_{int}={\bar \psi}{\hat \mu}\sigma^{{\hat a}{\hat
    b}}F_{{\hat a}{\hat b}}\psi\,,
 \]
where ${\hat \mu}$ is the magnetic momentum of the neutrino,
$F_{{\hat a}{\hat b}}$ is the electro-magnetic field tensor, and
$\sigma^{{\hat a}{\hat b}}=\frac{1}{4}[\gamma^{\hat a},
\gamma^{\hat b}]$. Spin flavor conversions of neutrinos
propagating in magnetic fields are generated by the transition
magnetic moments, which are non-diagonal terms in the effective
Hamiltonian describing the neutrino evolution.

Taking into account the gravitational and magnetic interactions,
the equation of evolution describing the conversion between two
neutrino flavors $f=e$ and $f'=\mu, \tau$ reads (we refer to Dirac
neutrinos, but the formalism also applies to Majorana neutrinos)
\cite{PIN}
 \begin{equation}\label{11}
i\frac{d}{d\lambda}\left(\begin{array}{c}
                           \nu_{fL} \\
                             \nu_{f' L} \\
                              \nu_{fR} \\
                           \nu_{f' R}\end{array}\right)={\cal H}\left(\begin{array}{c}
                                                            \nu_{fL} \\
                             \nu_{f' L} \\
                              \nu_{fR} \\
                           \nu_{f' R}\end{array}\right)\,,
 \end{equation}
where, in the chiral base, the matrix ${\cal H}$ is the effective
Hamiltonian
\begin{equation}\label{12}
{\cal H}=\left[\begin{array}{cc}
 {\cal H}_L & {\cal H}_{ff'}^* \vspace{0.05in} \\
 {\cal H}_{ff'} & {\cal H}_R \\
\end{array}\right]\,,
\end{equation}
\begin{equation}\label{12a}
{\cal H}_L=\left[\begin{array}{cc}
 V_{\nu_e}+\Omega_G-\delta c_2
         &  \delta s_2 \vspace{0.05in} \\
 \delta s_2  & V_{\nu_{f'}}+\Omega_G +\delta c_2 \\
\end{array}\right]\,,
\end{equation}
\begin{equation}\label{12b}
{\cal H}_R=\left[\begin{array}{cc}
 -\delta c_2 &  \delta s_2 \vspace{0.05in} \\
 \delta s_2  & \delta c_2 \\
\end{array}\right]\,, \quad
 {\cal H}_{ff'}=B_\perp \left[\begin{array}{cc}
 \mu_{ff} & \mu_{ff'} \vspace{0.05in} \\
 \mu_{ff'} & \mu_{f'f'} \\
\end{array}\right].
\end{equation}
\begin{eqnarray}\label{omegaG}
  \Omega_G(r)&=&\frac{p_\mu A^{\mu}_G(r)}{E_l}
  =\frac{4GMR^2}{5 r^3E_l}\, {\bf p}\cdot {\bbox \omega}' \\
 & \simeq &  8\,\, 10^{-13}
  \frac{M}{M_\odot}\left(\frac{R}{10\mbox{km}}\right)^2
  \left(\frac{10\mbox{km}}r \right)^3\frac{\omega' \cos \beta}{10^4\mbox{Hz}}
  \, \mbox{eV}\,. \nonumber
\end{eqnarray}
Here $\delta=\frac{\Delta m^2}{4E_l}$ ($\Delta m^2=m_2^2-m_1^2$),
$c_2=\cos 2\theta$, $s_2=\sin 2\theta$, $\theta$ is the vacuum
mixing angle, $E_l$ is the energy measured in the local frame,
$B_\perp=B\sin\alpha$ is the component of the magnetic field
orthogonal to the neutrino momentum, and $\beta$ is the angle
between the neutrino momentum and the angular velocity.

$\Omega_G$ is diagonal in spin space, so that it cannot induce
spin-flips. Its relevance comes from the fact that it modifies the
resonance conditions (resonances are governed by the $2\times 2$
submatrix in (\ref{12}) for each pairs of states). For the
transition $\nu_{fL}\to \nu_{f' R}$ one in fact gets
\begin{equation} \label{res1}
% \nu_{fL}\to \nu_{f' R} & \qquad \qquad&
V_{\nu_e}+\Omega_G({\bar r})- 2\delta c_2=0\,,
% \\   \nu_{f' L}\to \nu_{fR} & \quad \qquad &
% \Omega_B+\Omega_G({\bar r})+2\delta c_2=0\,. \label{res2}
\end{equation}
where ${\bar r}$ is the resonance point.

$\Omega_G$ in (\ref{res1}) distorts the surface of resonance owing
to the relative orientation of the neutrino momentum with respect
to the angular velocity. As a consequence, the outgoing energy
flux results modified. To estimate the anisotropy of the outgoing
neutrinos, one needs to evaluate the energy flux ${\bf F}_s$
emitted by nascent stars. According to the Barkovich, D'Olivo,
Montemayor and Zanella paper \cite{zanella}, the neutrino momentum
asymmetry is given by
\begin{equation}\label{asymmetrygeneral}
  \frac{\Delta p}{p}=\frac{1}{3}
  \frac{\int_0^\pi {\bf F}\cdot {\hat {\bbox \omega}}da}
  {\int_0^\pi {\bf F}\cdot {\hat {\bf n}}da}
    \sim -\frac{1}{9}\frac{\varrho}{\bar r}\,.
\end{equation}
The factor 1/3 comes from the fact that only one neutrinos species
is responsible for the anisotropy (thus it carries out only 1/3 of
the total energy). ${\bf F}_s$ is the outgoing neutrino flux
through the element area $da$ of the emission surface. ${\hat
{\bbox \omega}'}={\bbox \omega}'/|{\bbox \omega}'|$ is the
direction of the vector ${\bbox \omega'}$, whereas ${\bf {\hat
n}}$ is the unity vector orthogonal to $da$ (notice that the area
element in curved spacetime is $\sqrt{-g}\,da$; in the weak field
approximation one has $\sqrt{-g}\,da\sim da$). $\varrho$ is the
radial deformation of the effective surface of resonance. It
shifts the resonance point ${\bar r}$ to $r(\phi)={\bar r}+\varrho
\cos\phi$, with $\varrho\ll {\bar r}$ and $\cos \phi= {\hat{\bbox
\omega'}} \cdot {\bf {\hat p}}$. To determine ${\bar r}$, one uses
the resonance condition
 \begin{equation}\label{resV}
 2\delta c_2-V_{\nu_e}({\bar r})=0\,.
 \end{equation}

The deformation $\varrho$ is evaluated by expanding (\ref{res1})
about to $r(\phi)$, and using the shifts $p\to p+\delta_p$,
$V_{\nu_e}\to V_{\nu_e}+ \delta_{V_{\nu_e}}$ \cite{zanella}, where
 \begin{eqnarray}
 \delta_p&=&\frac{d\ln p}{dr}\,p\,\varrho\,\cos \phi=h_p^{-1}p\,\varrho\cos \phi\,,
 \label{deltap} \\
  \delta_{V_{\nu_e}}&=&\frac{d\ln V_{\nu_e}}{dr}\,V_{\nu_e}\,\varrho\,\cos\phi=
  h_{V_{\nu_e}}^{-1}\,V_{\nu_e}\,\varrho\cos \phi\,,
  \label{deltaV}
  \end{eqnarray}
Eqs. (\ref{resV}), (\ref{deltap}), and (\ref{deltaV}), allow to
write Eq. (\ref{asymmetrygeneral}) as
\begin{equation}\label{deltap1}
  \frac{\Delta p}{p}=\frac{4GMR^2\omega'}{5r^3}\frac{1}{9V_{\nu_e}{\bar r}(h_p^{-1}+h_{V_{\nu_f}}^{-1})}\,.
\end{equation}
To compute $h_p^{-1}+h_{V_{\nu_e}}^{-1}$, a model for the
protoneutron star has to be specified. For our purpose, we adopt
the polytropic model. The inner core of a protoneutron star is
consistently described by a polytropic gas of relativistic nucleon
with adiabatic index $\Gamma=4/3$ \cite{shapiro}. As in
\cite{zanella}, we assume that such a model also holds for the
rest of the star. The pressure $P$ and matter density $\rho$ are
related by $P=K\rho^\Gamma$ \cite{shapiro,zanella}, where
$K=T_c/m_n\rho_c^{1/3}\simeq 5.6\times 10^{-5}$MeV$^{-4/3}$.
$T_c=40$MeV and $\rho_c\simeq 10^{14}$gr/cm$^3$ are the
temperature and the matter density of the core, respectively. The
matter density profile $\rho(r)$ can be chosen as \cite{zanella}
\begin{equation}\label{rhoG}
 \rho^{\Gamma -1}(x)=\rho_c^{\Gamma-1}[a x^2+b x+c]\,,
\end{equation}
where $x=r_c/r$,
 \[
a=(1-\mu)\lambda_\Gamma\,, \quad b=(2\mu-1)\lambda_\Gamma\,, \quad
c=1-\mu\lambda_\Gamma\,,
\]
and $\lambda_\Gamma=G M_c (\Gamma-1)/r_c\rho_c^{\Gamma-1}K\Gamma
\simeq 0.87$. $r_c=10$km is the core radius and $M_c\simeq
M_\odot$ is the mass of the core ($M_\odot$ is the solar mass).
The parameter $\mu$ is determined by setting $\rho(R_s)=0$,
\begin{equation}\label{mu}
  \mu=\left[\frac{R_s}{\lambda_\Gamma(R_s-r_c)}-\frac{r_c}{R_s}\right]
  \frac{R_s}{R_s-r_c}\,.
\end{equation}
The temperature profile $T(r)$ is related to the matter density
$\rho(r)$ by the relation \cite{shapiro,zanella}
\begin{equation}\label{Tprofile}
  \frac{dT^2}{dr}=-\frac{9\kappa L_c}{\pi r^2}\, \rho\,,
\end{equation}
where $L_c\sim 9.5\times 10^{51}$erg/sec is the core luminosity,
and $\kappa\sim 6.2\times 10^{-56}$eV$^{-5}$. Eq. (\ref{Tprofile})
can be integrate via (\ref{rhoG})
\begin{equation}\label{T(r)}
  T(r)=T_c\, \sqrt{2\lambda_c[\chi(x)-\chi(1)]+1}\,,
\end{equation}
where $\lambda_c=9\kappa \, L_c \, \rho_c/2\pi T_c^2 r_c \sim
1.95$, and
\begin{equation}\label{chi}
  \chi(x)=c^{3}x+\frac{3}{2}\, b\,c^{2}x^2+c(a\,c+
  b^{2})\,x^3 +
\end{equation}
 \[
  \frac{b}{4}\,(6\,a\,c+b^{2})\,x^4+\frac{3\,a}{5}\,(a\,c+b^{2})\,x^5
  +\frac{b\,a^{2}}{2}\,x^6+\frac{a^{3}}{7}\,x^7\,.
 \]
Assuming the thermal equilibrium between neutrinos and the medium,
so that $p\sim T$ \cite{zanella}, and being $V_{\nu_e}\sim \rho$,
one can rewrite the inverse characteristic lengths $h_p^{-1}$ and
$h_{V_{\nu_e}}^{-1}$ as $h_p^{-1}\equiv h_T^{-1}$ and
$h_{V_{\nu_e}}^{-1}\equiv h_\rho^{-1}$. Eqs. (\ref{rhoG}) and
(\ref{T(r)}) imply (at the resonance)
\begin{eqnarray}
 h_T^{-1} &=& \frac{d\ln T}{dr} = -\lambda_c\, \frac{\rho({\bar r})}{\rho_c}\,
        \left(\frac{T_c}{T({\bar r})}\right)^2\,\frac{{\bar x}}{{\bar r}}\,,
        \label{hT} \\
 h_\rho^{-1} &=& \frac{d\ln \rho}{dr} = -3\, \left(
 \frac{\rho_c}{\rho({\bar r})} \right)^{1/3}\,
        \left(2\,a\, {\bar x}+b\right)\,\frac{{\bar x}}{{\bar r}}\,.  \label{hV}
\end{eqnarray}
As before pointed out, $V_{\nu_e}\ll 1$ ($y_e\ll 1$) at $r\sim
12$km. For resonances occurring at ${\bar r}\sim 12$km, where
$\rho({\bar r})\sim 10^{11}$gr/cm$^3$,
%Eq. (\ref{rhoG}) implies $\mu\sim 7.31$.
Eqs. (\ref{rhoG}) and (\ref{mu}) give $R_s\sim 12.5$km.  From Eqs.
(\ref{T(r)}), (\ref{hT}), (\ref{hV}) and (\ref{deltap1}), one
infers
\begin{equation}\label{yomega}
  \frac{1}{10^9 y_e}\,\frac{\omega}{10^4\mbox{Hz}}\sim 6\,,
\end{equation}
where $\Delta p/p\sim 0.01$ and $\omega'=2\omega$ (i.e. ${\bbox
\omega}\parallel {\bf x}$) have been used. Typical angular
velocities of pulsars $\omega\sim (1 - 10^4)$Hz lead to $y_e\sim
10^{-14} - 10^{-10}$, as follows from Eq. (\ref{yomega}).
%Fig. \ref{fig1} shows Eq. the resonance condition (\ref{resV}),
For neutrinos with momentum $p\sim 10$MeV, the resonance condition
(\ref{resV}) can be recast in the form $\Delta m^2 \cos 2 \theta
\simeq 7.6 \,\, 10^{4} y_e$eV$^2$, i.e. $10^{-10}\lesssim \Delta
m^2 \cos 2 \theta/\mbox{eV}^2 \lesssim 10^{-6}$. The latter is
consistent with the best fit of solar neutrinos $\Delta
m^2_{Sun}\sim (10^{-5}\div 10^{-4})\mbox{eV}^2$ and $0.8
\lesssim\sin^2 2 \theta_{Sun}\lesssim 1$ \cite{kamLAND}.

%\begin{figure}
%\resizebox{\hsize}{1.5in}{\includegraphics{Fig1.eps}} \caption{In
%figure it is represented $z_S$ vs $x_S$ and $y_S$.}
% \label{fig1}
%\end{figure}

Some comments are in order:

\begin{itemize}

\item In addition to the level crossing (\ref{res1}), it must also
occur that the crossing is adiabatic, i.e. the adiabatic parameter
$\gamma$, which quantifies the magnitude of the off-diagonal
elements with respect to the diagonal ones of (\ref{12}) in the
instantaneous eigenstates, must be greater than one when valuated
at the resonance ${\bar r}\sim 12$km. To show this, let us define
1) the precession length
 \[
L=\frac{2\pi}{\sqrt{(2\mu_{ff'}
B_\perp)^2+(V_{\nu_e}+\Omega_G-2\delta c_2)^2}}\,,
 \]
that at the resonance, it reduces to
 \[
 L_{res}=L({\bar r})=\frac{\pi}{\mu_{ff'} B_\perp}\simeq
 10^2 \,
 \frac{10^{-11}\mu_B}{\mu_{ff'}}\frac{10^{13}\mbox{G}}{B_\perp}\,\mbox{m}\,,
 \]
and 2) the width of the resonant spin flavor precession
 \[
\Delta_r=2 \lambda \Lambda\,,
 \]
being $\lambda = \pi/L_{res}\delta $ and
 \[
 \Lambda =\left(\frac{\rho'({\bar r})}{\rho({\bar r})}\right)^{-1}
 \approx\frac{V_{\nu_e}({\bar r})}{V_{\nu_e}'({\bar r})}\,.
 \]
$\Lambda$ is derived assuming $y_f'(r, t)=0$ ($Y_e'\ll
\rho'/\rho$) \cite{valle}. The spin flavor conversion is adiabatic
provided $\Delta_r \gg L_{res}$, or equivalently
 \[
 \frac{2(\mu_{ff'} B_\perp)^2}{\delta \pi |\rho'/\rho|}\equiv \gamma \gg
1\,,
 \]
where
 \[
\mu_{ff'}B_\perp\sim 5.6\times
10^{-7}\frac{\mu_{ff'}}{10^{-11}\mu_B}\frac{B_\perp}{10^{13}\mbox{G}}\,
\mbox{eV}\,.
 \]
Since the magnetic fields inside to protoneutron stars are greater
than $B\gtrsim 10^{11}$G, whereas $\mu_{ff'}\lesssim
10^{-11}\mu_B$, as provided by astrophysical and cosmological
constraints \cite{pakvasa,raffelt}, one gets $\gamma \gg 1$, i.e.
the adiabatic condition is fulfilled (such a result follows by
using Eq. (\ref{rhoG})).

\item  The conversion probability $P_{\nu_{fL}\to \nu_{f'
R}}$ is given by \cite{okun}
 \[
 P_{\nu_{fL}\to \nu_{f' R}}=1/2-(1/2-P)\cos
 2{\tilde \theta}_i \cos 2 {\tilde \theta}_f\,,
 \]
where $P=e^{-\gamma \pi/2}$ is the Landau-Zener probability, and
${\tilde \theta}$ is defined as
 \[
 \tan 2 {\tilde
 \theta}(r)=2\mu_{ff'}B_\perp/(\Omega_G(r)+V_{\nu_e}-2\delta c_2)\,.
 \]
${\tilde \theta}_i={\tilde \theta}(r_i)$ is the initial mixing
angle of neutrinos produced at $r_i$, and ${\tilde
\theta}_f={\tilde \theta}(r_f)$ is the mixing angle of neutrinos
where the spin flip probability is evaluated. Since $\gamma \gg
1$, the Landau-Zener probability $P$ vanishes.

\item The weak field approximation is fulfilled
since $(4GMR^2/5{\bar r}^3)\omega R_s\lesssim 10^{-2}$ as
$\omega\lesssim 10^4$Hz and $R_s\gtrsim 12$ km. This means that
the geometry of the rotating sources is correctly described by
(\ref{9}) up to angular velocities $\omega \sim 10^4$Hz. Since the
typical angular velocity of the protoneutron star varies in the
range $(10^2-10^3)$Hz, the weak field approximation here used can
be applied.

\item The rotational effects are relevant during the time
scale $t_0\lesssim 10$ sec ($t_0$ is the time scale for the
emission of the energy $\sim 0.5\times 10^{53}$erg by each
neutrinos degree of freedom with $p\sim 10$MeV) \cite{raffelt}.

\end{itemize}

\section{Conclusion}

In this paper it has been suggested a mechanism for the generation
of pulsar kicks which accounts for the spin-gravity coupling of
neutrinos propagating in a gravitational field of a rotating
nascent star. Owing to the relative orientation of neutrino
momenta with respect to the direction of the angular velocity, the
energy flux turns out to be generated anisotropically. Results
imply a correlation between the motion of pulsars and their
angular velocities. Such a connection seems to be corroborated by
recent statistical analysis and observations discussed in
\cite{laiOmega,wex}.

Spin-gravity coupling is strictly related to the so called
gravito-magnetic effect, an effect predicted by General Relativity
\cite{ciufolini}, as well as by many metric theories
\cite{will,barros,camacho}. Its origin is due to the mass-energy
currents (moving or rotating matter contributes to the
gravitational field, in analogy to the magnetic field of moving
charges or magnetic dipole). Experiments involving the technology
of laser ranged satellites \cite{ciufoliniScience,ciufoliniPLA}
are at the moment the favorite candidate to test gravito-magnetic
effects. In connections with the mechanism proposed in this paper,
a direct evidence of the gravito-magnetic effect might be provided
by pulsar kicks. Future investigations on the velocity
distribution of pulsars will certainly allow to clarify this still
open issue.

Results of this paper (as well as the papers
\cite{burrows,cuesta,loveridge}, in which pulsar kicks are
discussed in relation to gravitational waves) have been obtained
in semiclassical approximation, i.e. the gravitational field is
described by the classical field equations of General Relativity.
It will be of interest to investigate within the framework of
quantum gravity theories.

%\vspace{0.1in}

{\it Acknowledgments}: It is a pleasure to thank A. Kusenko and
J.F. Nieves, for the illumining discussions. Thanks to Mosquera
Cuesta and P.B. Pal for their comments. Many thanks to M.
Barkovich, J.C. D'Olivo and R. Montemayor for elucidations on
topics related to neutrino emissions.

%\acknowledgments Research supported by MURST PRIN 2003.

\end{document}